\documentclass[9pt,conference]{IEEEtran}
\IEEEoverridecommandlockouts
\usepackage{babel}
\usepackage[T1]{fontenc}
\usepackage{cite}
\usepackage{amsmath,amssymb,amsfonts}
\usepackage{amsfonts,mathtools}
\usepackage{empheq}
\usepackage{bm,bbm,dsfont}
\usepackage{algorithmic}
\usepackage{graphicx}
\usepackage{subcaption}
\usepackage{textcomp}
\usepackage{xcolor}
\usepackage{pgfplots}
\usepackage{tikz}
\pgfplotsset{compat=newest}
\pgfplotsset{plot coordinates/math parser=false} 
\usepgfplotslibrary{patchplots}
\usepgfplotslibrary{colormaps}%
\def\BibTeX{{\rm B\kern-.05em{\sc i\kern-.025em b}\kern-.08em
    T\kern-.1667em\lower.7ex\hbox{E}\kern-.125emX}}
\usepackage{microtype}
\usepackage[subtle,title=normal,sections=normal,margins=normal,bibliography=normal]{savetrees}
\usepackage[bookmarksopen=false,hidelinks]{hyperref}
\DeclarePairedDelimiter\curlyBrace\{\}
\newcommand{\myExpectation}[1]{\operatorname{\mathcal{E}}\curlyBrace*{#1}}
\DeclarePairedDelimiter\abs{\lvert}{\rvert}%
\DeclarePairedDelimiter\norm{\lVert}{\rVert}%
\newcommand{\dom}{\mathrm{d}}
\newcommand{\thetaDom}{\theta_{\dom}}
\newcommand{\head}{\mathrm{HA}}
\newcommand{\eMicInd}{\mathrm{E}}
\newcommand{\vectorADom}{\mathbf{a}\left(\thetaDom\right)}
\newcommand{\vectorADomHerm}{\mathbf{a}^{H}\left(\thetaDom\right)}
\newcommand{\vectorADomwhite}{\mathbf{a}^{\mathrm{w}}\left(\thetaDom\right)}
\newcommand{\vectorADomHermwhite}{\left(\mathbf{a}^{\mathrm{w}}\left(\thetaDom\right)\right)^{H}}
\newcommand{\vectorADomwhiteCMA}{\mathbf{a}_{\head}^{\mathrm{w}}\left(\thetaDom\right)}
\newcommand{\vectorADomHermwhiteCMA}{\left(\mathbf{a}_{\head}^{\mathrm{w}}\left(\thetaDom\right)\right)^{H}}
\newcommand{\vectorGDom}{\mathbf{g}\left(\thetaDom\right)}
\newcommand{\vectorGhat}{\hat{\mathbf{g}}\left(k,l\right)}
\newcommand{\vectorGproto}{\bar{\mathbf{g}}\left(k,\theta_{i}\right)}
\newcommand{\vectorGprotoHerm}{\bar{\mathbf{g}}^{H}\left(k,\theta_{i}\right)}
\newcommand{\vectorXDom}{\mathbf{x}_{\dom}}
\newcommand{\vectorXDomDP}{\mathbf{x}_{\dom}^{\mathrm{DP}}}
\newcommand{\vectorXDomRev}{\mathbf{x}_{\dom}^{\mathrm{R}}}
\newcommand{\vectorN}{\mathbf{n}}
\newcommand{\vectorU}{\mathbf{u}}
\newcommand{\vectorUherm}{\mathbf{u}^{H}}
\newcommand{\vectorY}{\mathbf{y}}
\newcommand{\vectorYherm}{\mathbf{y}^{H}}
\newcommand{\phiY}{\boldsymbol{\Phi}_{\mathrm{y}}}
\newcommand{\phiYHA}{\boldsymbol{\Phi}_{\mathrm{y,\head}}}
\newcommand{\phiYwhite}{\boldsymbol{\Phi}_{\mathrm{y}}^{\mathrm{w}}}
\newcommand{\phiYwhiteHA}{\boldsymbol{\Phi}_{\mathrm{y,\head}}^{\mathrm{w}}}
\newcommand{\phiXDomDP}{\boldsymbol{\Phi}_{\mathrm{x}_{\dom}}^{\rm{DP}}}
\newcommand{\phiU}{\boldsymbol{\Phi}_{\mathrm{u}}}
\newcommand{\phiUHA}{\boldsymbol{\Phi}_{\mathrm{u,\head}}}
\newcommand{\selectionMatrixHA}{\mathbf{E}_{\head}}
\newcommand{\hatphiY}{\hat{\boldsymbol{\Phi}}_{\mathrm{y}}}
\newcommand{\hatphiYwhite}{\hat{\boldsymbol{\Phi}}_{\mathrm{y}}^{\mathrm{w}}}
\newcommand{\hatphiYwhiteHA}{\hat{\boldsymbol{\Phi}}_{\mathrm{y,\head}}^{\mathrm{w}}}
\newcommand{\hatphiU}{\hat{\boldsymbol{\Phi}}_{\mathrm{u}}}

\newcommand{\myIdentity}{\mathbf{I}}%
\newcommand{\PSDs}{\phi_{\mathrm{s}}}%
\newcommand{\indicatorFunctionKL}{\mathbbm{1}_{j}\left(k,l\right)}
\begin{document}

\title{Completing Sets of Prototype Transfer Functions for Subspace-based Direction of Arrival Estimation of Multiple Speakers
\thanks{This work was funded by the Deutsche Forschungsgemeinschaft (DFG, German Research Foundation) under Germany's Excellence Strategy - EXC 2177/1 - Project ID 390895286 and Project ID 352015383 - SFB 1330 B2.\newline ©20XX IEEE.  Personal use of this material is permitted. Permission from IEEE must be obtained for all other uses, in any current or future media, including reprinting/republishing this material for advertising or promotional purposes, creating new collective works, for resale or redistribution to servers or lists, or reuse of any copyrighted component of this work in other works.}
}

\author{\IEEEauthorblockN{Daniel Fejgin and Simon Doclo}
\IEEEauthorblockA{Dept. of Medical Physics and Acoustics and Cluster of Excellence Hearing4all,\\ Carl von Ossietzky Universität Oldenburg, Germany}
}

\maketitle

\begin{abstract}
To estimate the direction of arrival (DOA) of multiple speakers, subspace-based prototype transfer function matching methods such as multiple signal classification (MUSIC) or relative transfer function (RTF) vector matching are commonly employed. In general, these methods require calibrated microphone arrays, which are characterized by a known array geometry or a set of known prototype transfer functions for several directions. In this paper, we consider a partially calibrated microphone array, composed of a calibrated binaural hearing aid and a (non-calibrated) external microphone at an unknown location with no available set of prototype transfer functions. We propose a procedure for completing sets of prototype transfer functions by exploiting the orthogonality of subspaces, allowing to apply matching-based DOA estimation methods with partially calibrated microphone arrays. For the MUSIC and RTF vector matching methods, experimental results for two speakers in noisy and reverberant environments clearly demonstrate that for all locations of the external microphone DOAs can be estimated more accurately with completed sets of prototype transfer functions than with incomplete sets.
\end{abstract}

\begin{IEEEkeywords}
direction of arrival estimation, subspaces, binaural hearing aids, external microphone 
\end{IEEEkeywords}

\section{Introduction}
\label{sec:intro}
Many speech communication applications such as hearing aids and hands-free conferencing systems require estimates of the direction of arrival (DOA) of multiple speakers in noisy and reverberant environments. Over the last decades, many model-based and machine-learning-based methods have been developed for DOA estimation, typically requiring knowledge about the geometrical configuration of the microphone array \cite{Cobos2017,Gannot2019,Evers2020,Grumiaux2022,Schmidt19886,DiBiase2001,Tourbabin2015,Chakrabarty2019,Hafezi2019,Hammer2021}. However, applying these methods to DOA estimation using microphone arrays with a partially unknown geometrical configuration, e.g., a hearing aid linked to one or more external microphones (eMics) at unknown locations, is not straightforward.

In \cite{Farmani2018,Fejgin2023FA,Kowalk2022,Fejgin2021,Bruemann2024,Fejgin2023} DOA estimation methods have been proposed which exploit the availability of one or more eMics. The methods in \cite{Farmani2018} and \cite{Kowalk2022} utilize the eMic signal to estimate a clean speech reference signal and a voice activity detector, however, severely restricting the location of the eMic to the vicinity of the target speaker. Without restricting the location of the eMic, the methods in \cite{Fejgin2021,Fejgin2023FA}, and \cite{Bruemann2024} utilize the eMic signal to estimate relative transfer function (RTF) vectors or generalized cross correlations which are subsequently used to construct a spatial spectrum. However, since the location of the eMic is unknown only a spatial spectrum for the microphones of the calibrated array instead of a spatial spectrum for all microphones is constructed. In this context, calibrated arrays are defined as arrays for which a set of (anechoic) prototype transfer function vectors for several directions is available, either because the array configuration is known or a set of measured transfer functions is available. Opposed to these methods, the method in \cite{Fejgin2023} utilizes the signals from a calibrated array of eMics at unknown locations to estimate RTF vectors as well as to construct a spatial spectrum for all available microphones.

To perform DOA estimation using prototype transfer function vector matching-based methods, in this paper we aim at exploiting the eMic for the construction of a spatial spectrum for all available microphones, however, without requiring the availability of prototype transfer functions for the eMic. We propose an optimal procedure in the least-squares sense that completes sets of prototype transfer function vectors with a transfer function corresponding to the eMic. We complete these sets by exploiting the orthogonality of the complementary signal subspace and the noise subspace obtained from the eigenvalue decomposition of the covariance matrix of pre-whitened signals. For a binaural hearing aid setup with an external microphone at an unknown location, we compare the DOA estimation performance with the incomplete set (i.e., using only the microphone signals of the hearing aids) and the proposed completed sets of prototype transfer function vectors using MUSIC \cite{Schmidt19886} and the RTF vector matching method \cite{Fejgin2021}. Results using real-world recordings from the BRUDEX database \cite{Fejgin2023ITG} clearly demonstrate a benefit of the proposed method for multiple eMic locations that are distributed over a large area.

\section{Signal model and notation}
\label{sec:sigModel}
In a noisy and reverberant acoustic environment, we consider $J$ speakers which are recorded using a binaural hearing aid setup with $M/2$ microphones on each hearing aid and one eMic, i.e., a total of $M +1$ microphones. The eMic is spatially separated from the hearing aids at an unknown location (see Fig. 1). In the short-time Fourier transform (STFT) domain, the $m$-th microphone signal can be written as
\begin{equation}
	\label{eq:signalModel_micComponent}
	Y_{m}\left(k,l\right) = \sum_{j=1}^{J}X_{m,j}\left(k,l\right) + N_{m}\left(k,l\right)\,,
\end{equation}
where $X_{m,j}$ and $N_{m}$ denote the speech component of the \mbox{$j$-th} speaker and the noise component, respectively, where $m\in\left\{1,\dots,M+1=\eMicInd\right\}$, $k\in\left\{1,\dotsc,K\right\}$ and $l\in\left\{1,\dotsc,L\right\}$ denote the microphone index, the frequency bin index, and the frame index, res\-pec\-ti\-ve\-ly. Denoting the stacked vector of microphone signals as $\vectorY\left(k,l\right) = \left[Y_{1}\left(k,l\right),\,\dots,\,Y_{\eMicInd}\left(k,l\right)\right]^{T}\in\mathbb{C}^{M+1}$, with $\left(\cdot\right)^{T}$ denoting the transposition operator and assuming disjoint speaker activity in the STFT domain \cite{Yilmaz2004}, this vector can be approximated as $\vectorY\left(k,l\right)\approx\vectorXDom\left(k,l\right) + \vectorN\left(k,l\right)$, where the vectors $\vectorXDom$ and $\vectorN$ denote the speech component of the dominant speaker and the noise component, respectively, both defined similarly as $\vectorY$. For conciseness, we omit the time-frequency (TF) bin indices $k$ and $l$ when possible in the following.

We assume that the speech component $\vectorXDom$ can be split into a direct-path component $\vectorXDomDP$ and a reverberant component $\vectorXDomRev$, i.e., $\vectorXDom = \vectorXDomDP + \vectorXDomRev$. Condensing the noise and reverberation components into the undesired component $\vectorU = \vectorN + \vectorXDomRev$, the vector of microphone signals can be written as $\vectorY = \vectorXDomDP + \vectorU$. 

Approximating the direct-path component with a multiplicative transfer function \cite{Avargel2007} using the direct-path acoustic transfer function (ATF) vector $\vectorADom = \left[A_{1}\left(\thetaDom\right),\,\dots,\,A_{\eMicInd}\left(\thetaDom\right)\right]^{T}$, with $\thetaDom$ denoting the DOA of the dominant speaker relative to the binaural hearing aid setup, the vector $\vectorXDomDP$ can be written as
\begin{equation}
	\label{eq:rank1Assumption}
	\vectorXDomDP = \vectorADom S_{\dom} = \vectorGDom X_{1,\mathrm{d}}\,,
\end{equation}
where $S_{\mathrm{d}}$ denotes the dominant speech signal and $\vectorGDom = \vectorADom/A_{1}\left(\thetaDom\right)$ denotes the direct-path relative transfer function (RTF) vector, assuming the first microphone to be the reference microphone. For the binaural hearing aid setup only, we assume the availability of a set of anechoic prototype ATF vectors $\bar{\mathbf{a}}_{\head}\left(k,\theta_{i}\right)$ for different candidate directions $\theta_{i}\,,i=1,\dotsc,I$. The set of anechoic prototype RTF vectors is obtained as $\bar{\mathbf{g}}_{\head}\left(k,\theta_{i}\right) = \frac{\bar{\mathbf{a}}_{\head}\left(k,\theta_{i}\right)}{\mathbf{e}_{1}^{T}\bar{\mathbf{a}}_{\head}\left(k,\theta_{i}\right)}$, where $\mathbf{e}_{m} = [0,...,1,0,...,0]$ denotes a selection vector with all zeros except the $m$-th element. Since the location of the eMic is unknown, obviously no sets of anechoic prototype ATFs $\bar{A}_{\eMicInd}\left(k,\theta_{i}\right)$ and RTFs $\bar{G}_{\eMicInd}\left(k,\theta_{i}\right)$ are available for the eMic. Therefore, the considered microphone array is referred to as partially calibrated.

Assuming uncorrelated direct-path speech and undesired components, the $\left(M+1\right)\times \left(M+1\right)$ noisy covariance matrix can be written, using \eqref{eq:rank1Assumption}, as
\begin{equation}
	\label{eq:covMats}
	\phiY = \myExpectation{\vectorY\vectorYherm} = \phiXDomDP+\phiU = \PSDs\vectorADom\vectorADomHerm+\phiU\,,
\end{equation}
where $\left(\cdot\right)^{H}$ and $\myExpectation{\cdot}$ denote the complex transposition and expec\-ta\-tion operators, respectively, $\phiU = \myExpectation{\vectorU\vectorUherm}$ denotes the covariance matrix of the undesired component, and $\PSDs = \mathcal{E}\{\big|S_{\dom}\big|^{2}\}$ denotes the power spectral density of the dominant speaker. Based on \eqref{eq:covMats} and a square-root decomposition (e.g., Cholesky decomposition) of the covariance matrix of the undesired component $\phiU = \phiU^{1/2}\phiU^{H/2}$, the noisy covariance matrix after pre-whitening can be written as
\begin{equation}
		\label{eq:covMatsWhite}
		\phiYwhite = \phiU^{-1/2}\phiY\phiU^{-H/2} =
		\PSDs\vectorADomwhite\vectorADomHermwhite  + \myIdentity_{M+1\times M+1}\,,
\end{equation}
with $\vectorADomwhite = \phiU^{-1/2}\vectorADom$ denoting the pre-whitened direct-path ATF vector. The $M\times M$ covariance matrices $\phiYHA$, $\phiYwhiteHA$, and $\phiUHA$ corresponding to the hearing aid signals can be extracted from \eqref{eq:covMats} and \eqref{eq:covMatsWhite} as $\phiYHA = \selectionMatrixHA\phiY\selectionMatrixHA^{T}$, $\phiYwhiteHA = \selectionMatrixHA\phiYwhite\selectionMatrixHA^{T}$ and $\phiUHA = \selectionMatrixHA\phiU\selectionMatrixHA^{T}$ with the selection matrix $\selectionMatrixHA = \left[\myIdentity_{M\times M},\mathbf{0}_{M\times 1}\right]$, where $\myIdentity_{M\times M}$ denotes an $M\times M$ identity matrix and $\mathbf{0}_{M\times 1}$ denotes a vector with zeros.

\section{Subspace-based DOA Estimation Methods}
\label{sec:doaEst}
\begin{figure*}
\scalebox{0.91}{
\begin{subfigure}{.47\textwidth}
	\centering
	\includegraphics[width=0.5\linewidth,trim={2.15cm 12.5cm 2.25cm 6.5cm},clip]{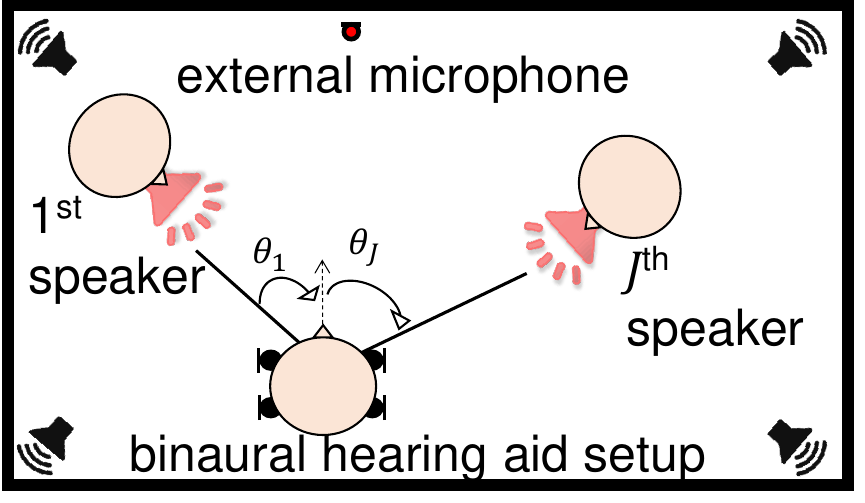}
	\caption{Fig. 1: Considered noisy and reverberant acoustic scenario with $J$ speakers, a binaural hearing aid setup and an external microphone.\label{fig:acousticScenario}}
\end{subfigure}
}
\hskip1.2cm
\scalebox{0.91}{
\begin{subfigure}{.47\textwidth}
	\centering
	\begin{tikzpicture}[baseline=(current bounding box.north)]
		\node[anchor=north,inner sep=0] at (0,0) {\includegraphics[width=0.95\linewidth,trim={3cm 8cm 3cm 4cm},clip]{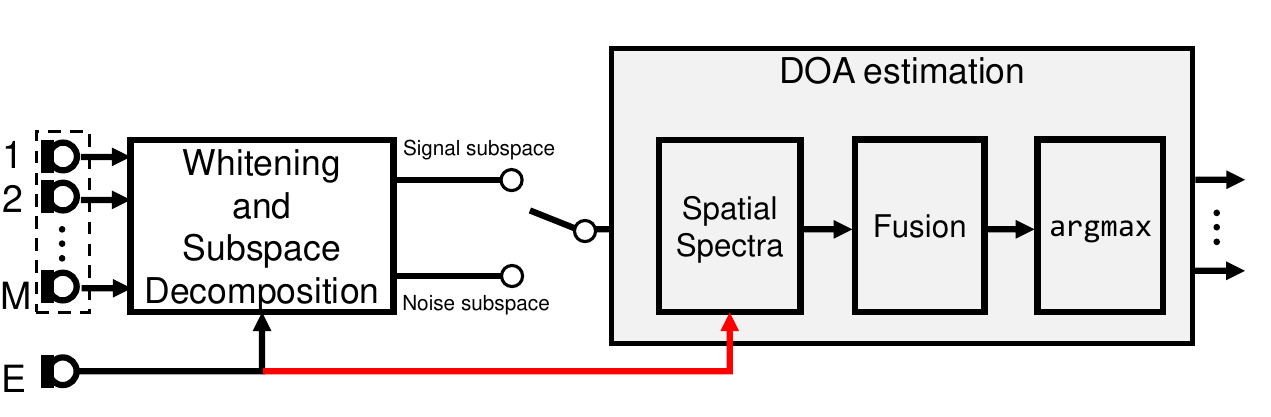}};
		\node[anchor=north,inner sep=0, xshift=4cm,yshift=-0.65cm,font = \fontsize{10}{1}\selectfont,]{$\hat{\theta}_{1}\left(l\right)$};
		\node[anchor=north,inner sep=0, xshift=4cm,yshift=-1.85cm,font = \fontsize{10}{1}\selectfont,]{$\hat{\theta}_{J}\left(l\right)$};
	\end{tikzpicture}
	\caption{Fig. 2: Block diagram of the baseline (without red arrow) and proposed (with arrow) subspace-based DOA estimation methods\label{fig:blockDiagram}.}
\end{subfigure}
}
\end{figure*}
To estimate the DOAs $\theta_{1:J}$ of all speakers, we consider two baseline methods, namely MUSIC \cite{Schmidt19886} and the RTF vector matching method presented in \cite{Fejgin2021}. Both methods consider complementary subspaces obtained from the same subspace decomposition. In Section \ref{ssec:doaEst__SPS}, we review for both methods the construction of frequency-dependent spatial spectra (SPS), i.e., functions of candidate directions with the location of peaks likely corresponding to the true speaker DOAs. In Section \ref{ssec:doaEst__fusion}, we review how DOAs are estimated from these frequency-dependent SPS (see Fig. 2).

\subsection{Construction of frequency-dependent SPS}
\label{ssec:doaEst__SPS}
The frequency-dependent SPS for both baseline methods are obtained from the eigenvalue decomposition (EVD) of an estimate of the noisy covariance matrix $\hatphiY$ after pre-whitening with a square-root decomposition (e.g., Cholesky decomposition) of an estimate of the covariance matrix of the undesired component $\hatphiU = \hatphiU^{1/2}\hatphiU^{H/2}$ \cite{Schmidt19886,Swindlehurst1992,Fejgin2021}, i.e.
\begin{equation}
	\hatphiYwhite = \hatphiU^{-1/2}\hatphiY\hatphiU^{-H/2} \overset{\mathrm{EVD}}{=}\hat{\mathbf{Q}}\hat{\boldsymbol{\Lambda}}\hat{\mathbf{Q}}^{H}\,,
\end{equation}
where $\hat{\mathbf{Q}}$ and $\hat{\boldsymbol{\Lambda}}$ denote the unitary matrix containing the eigenvectors and the diagonal matrix containing the eigenvalues of the pre-whitened matrix $\hatphiYwhite$, respectively. The matrix of eigenvectors can be partitioned into the signal subspace $\mathcal{P}\{\hatphiYwhite\}$ and noise subspace $\hat{\mathbf{Q}}_{\mathrm{n}}$ as $\hat{\mathbf{Q}} = \begin{bmatrix}
	\mathcal{P}\{\hatphiYwhite\},\hspace{2pt}\hat{\mathbf{Q}}_{\mathrm{n}}
\end{bmatrix}$, where $\mathcal{P}\left\{\cdot\right\}$ denotes the principal eigenvector of a matrix. 

The signal subspace of $\hatphiYwhite$ is spanned by the principal eigenvector $\mathcal{P}\{\hatphiYwhite\}$ and based on \eqref{eq:covMatsWhite} it is assumed that this vector does not differ too much from the vector $\vectorADomwhite$. This assumption is utilized by the state-of-the-art covariance whitening (CW) method \cite{Markovich2009} to estimate RTF vectors via de-whitening and normalization, i.e.,
\begin{equation}
	\label{eq:CW}
	\vectorGhat = \frac{\hatphiU^{1/2}\left(k,l\right)\mathcal{P}\{\hatphiYwhite\left(k,l\right)\}}{\mathbf{e}_{1}^{T}\hatphiU^{1/2}\left(k,l\right)\mathcal{P}\{\hatphiYwhite\left(k,l\right)\}}\,.
\end{equation}

The noise subspace of $\hatphiYwhite$ is spanned by the columns of $\hat{\mathbf{Q}}_{\mathrm{n}}$. Both subspaces are orthogonal to each other. For the construction of frequency-dependent SPS, MUSIC considers only the noise subspace whereas the RTF vector matching method considers only the signal subspace.

Exploiting the orthogonality between the signal and noise subspaces, MUSIC estimates the speaker DOAs by searching for vectors from a set of pre-whitened prototype ATF vectors $\bar{\mathbf{a}}^{\mathrm{w}}\left(k,l,\theta_{i}\right) = \hatphiU^{-1/2}\left(k,l\right)\bar{\mathbf{a}}\left(k,\theta_{i}\right) = [\bar{\mathbf{a}}^{\mathrm{w}}_{\head}\left(k,l,\theta_{i}\right),\,\bar{A}^{\mathrm{w}}_{\eMicInd}\left(k,l,\theta_{i}\right)]^{T}$ with $\bar{\mathbf{a}}\left(k,\theta_{i}\right) = [\bar{\mathbf{a}}_{\head}\left(k,\theta_{i}\right),\,\bar{A}_{\eMicInd}\left(k,\theta_{i}\right)]^{T}$ for different candidate directions $\theta_{i}$ that maximize the ortho\-go\-na\-lity with the estimated noise subspace. Thus, the frequency-dependent SPS is constructed as follows
\begin{equation}
	\label{eq:MUSICrawschmidt}
	\tilde{p}^{\mathrm{MUSIC}}\left(k,l,\theta_{i}\right) =
	\frac{1}{\norm{\hat{\mathbf{Q}}_{\mathrm{n}}^{H}\left(k,l\right)\bar{\mathbf{a}}^{\mathrm{w}}\left(k,l,\theta_{i}\right)}_{2}^{2}}\,.
\end{equation}
Instead of considering the SPS in \eqref{eq:MUSICrawschmidt} directly for DOA estimation, we will consider the normalized SPS as suggested by \cite{Salvati2014}, i.e.,
\begin{equation}
	\label{eq:MUSIC}
	p^{\mathrm{MUSIC}}\left(k,l,\theta_{i}\right) = \frac{\tilde{p}^{\mathrm{MUSIC}}\left(k,l,\theta_{i}\right)}{\underset{\theta_{i^{\prime}}}{\mathrm{max}}\hspace{2pt}\tilde{p}^{\mathrm{MUSIC}}\left(k,l,\theta_{i^{\prime}}\right)}\,.
\end{equation}

Exploiting the assumed parallelity of the estimated signal subspace with the pre-whitened direct-path ATF vector, the RTF vector matching method estimates the speaker DOAs by searching for vectors from a set of prototype RTF vectors $\bar{\mathbf{g}}\left(k,\theta_{i}\right) = [\bar{\mathbf{g}}_{\head}\left(k,\theta_{i}\right),\,\bar{G}_{\eMicInd}\left(k,\theta_{i}\right)]^{T}$ for different candidate directions $\theta_{i}$ that maximize the parallelity with the estimated RTF vectors. Considering the Hermitian angle \cite{Scharnhorst2001} between the estimated and prototype RTF vector as a measure for parallelity, the frequency-dependent SPS is constructed as follows
\begin{equation}
	\label{eq:RTFprototypeMatching}
	p^{\mathrm{RTF}}\left(k,l,\theta_{i}\right) = - \mathrm{arccos}\left(\frac{\abs{\vectorGprotoHerm\vectorGhat}}{\norm{\vectorGproto}_{2}\norm{\vectorGhat}_{2}}\right)\,.
\end{equation}

It should be noted that in order to compute the SPS according to \eqref{eq:MUSIC} and \eqref{eq:RTFprototypeMatching}, prototype transfer functions $\bar{\mathbf{a}}^{\mathrm{w}}\left(k,l,\theta_{i}\right)$ and $\bar{\mathbf{g}}\left(k,\theta_{i}\right)$ for \textit{all} microphones must be available. Since for the considered partially calibrated microphone array only $\bar{\mathbf{a}}_{\head}^{\mathrm{w}}\left(k,l,\theta_{i}\right)$ and $\bar{\mathbf{g}}_{\head}\left(k,\theta_{i}\right)$ are available, these SPS cannot be calculated. Hence, to estimate DOAs using the considered methods one can either consider the signals of the binaural hearing aid setup only (leading to the construction of SPS obtained from the subspace decomposition of $\hatphiYwhiteHA = \selectionMatrixHA\hatphiYwhite\selectionMatrixHA^{T}$) or one needs to complete the sets of prototype transfer function vectors $\bar{\mathbf{a}}_{\head}^{\mathrm{w}}\left(k,l,\theta_{i}\right)$ and $\bar{\mathbf{g}}_{\head}\left(k,\theta_{i}\right)$ with elements corresponding to the eMic, which is the topic of this paper (see Section \ref{sec:mainProp}).

\subsection{Fusion of frequency-dependent SPS}
\label{ssec:doaEst__fusion}
To exploit spatial information across multiple frequencies, the spatial spectra $p\left(k,l,\theta_{i}\right)$ in \eqref{eq:MUSIC} or \eqref{eq:RTFprototypeMatching} are combined as proposed in \cite{Fejgin2024} using the speaker-grouped frequency fusion mechanism \cite{Blandin2012}. Assuming the number of speakers $J$ to be known, the speaker DOAs are estimated from frequency-averaged SPS, each associated with a single speaker, as
\begin{equation}
	\hat{\theta}_{j}\left(l\right) = \underset{\theta_{i}}{\mathrm{argmax}}\sum_{k\in\mathcal{K}\left(l\right)}\hspace{-2pt}\indicatorFunctionKL p\left(k,l,\theta_{i}\right)\,, \quad j=1,...,J\,,
\end{equation}
where $\mathcal{K}$ denotes the subset of frequencies where one speaker is assumed to dominate over all other speakers, noise, and reverberation, and $\indicatorFunctionKL$ denotes an indicator function that denotes the association between the $\left(k,l\right)$-th TF bin and the $j$-th speaker, i.e.,
\begin{equation}
	\indicatorFunctionKL = \begin{cases}
		1 & \text{TF bin }\left(k,l\right)\text{ associated with speaker $j$}\\
		0 & \text{else}\,.
	\end{cases}
	\label{eq:TFassociation_general}
\end{equation}

As proposed in \cite{Fejgin2022}, to estimate the frequency subset $\mathcal{K}\left(l\right)$ we consider a criterion based on the binaural effective-coherence-based coherent-to-diffuse ratio (CDR), i.e.,
\begin{equation}
	\label{eq:CDRcriterion}
	\mathcal{K}\left(l\right) = \left\{k: \widehat{\mathrm{CDR}}\left(k,l\right)\geq\mathrm{CDR}_{\mathrm{thresh}}\right\}\,,
\end{equation}
where $\widehat{\mathrm{CDR}}$ denotes an estimate of the CDR and $\mathrm{CDR}_{\mathrm{thresh}}$ denotes a threshold value. As proposed in \cite{Fejgin2024}, to estimate the indicator function $\indicatorFunctionKL$ we dis\-cri\-mi\-nate the speakers spatially using estimated interaural time differences $\hat{\tau}_{j}\left(l\right)$ and a score function $\Psi\left(k,l,\hat{\tau}_{j}\left(l\right)\right)$, i.e.
\begin{equation}
	\indicatorFunctionKL = \begin{cases}
		1 & \text{if } j= \underset{j^{\prime}\in\left\{1,...,J\right\}}{\mathrm{argmax}}\,\Psi\left(k,l,\hat{\tau}_{j^{\prime}}\left(l\right)\right)\\
		0 & \text{else}\,.
	\end{cases}
	\label{eq:TFassociation_specific}
\end{equation}
Details regarding the estimation of $\mathcal{K}$ and $\indicatorFunctionKL$ can be found in \cite{Fejgin2022} and \cite{Fejgin2024}, respectively.

\section{Completing Sets of Prototype Transfer Functions}
\label{sec:mainProp}
In order to construct frequency-dependent SPS according to \eqref{eq:MUSIC} and \eqref{eq:RTFprototypeMatching}, in this section we propose a procedure that completes sets of prototype transfer function vectors.

Performing an EVD on the noisy covariance matrix after pre-whitening $\phiYwhite$ in \eqref{eq:covMatsWhite}, results in $\phiYwhite = \mathbf{Q}\boldsymbol{\Lambda}\mathbf{Q}^{H}$, where $\mathbf{Q} = \begin{bmatrix}
	\vectorADomwhite,\hspace{2pt}\mathbf{Q}_{\mathrm{n}}
\end{bmatrix}$ and $\mathbf{Q}_{\mathrm{n}}$ denotes the noise subspace of $\phiYwhite$. Partitioning the noise subspace as $\mathbf{Q}_{\mathrm{n}} = [\mathbf{Q}_{\mathrm{n},\head}^{T},\hspace{2pt}\mathbf{q}_{\mathrm{n},\eMicInd}]^{T}$ with $\mathbf{Q}_{\mathrm{n},\head} = \selectionMatrixHA\mathbf{Q}_{\mathrm{n}}$ and $\mathbf{q}_{\mathrm{n},\eMicInd} = \mathbf{Q}_{\mathrm{n}}^{T}\mathbf{e}_{\eMicInd}$, allows to write the orthogonality of the signal subspace with the noise subspace as
\begin{align}
	\mathbf{Q}_{\mathrm{n}}^{H}\vectorADomwhite = 
	\begin{bmatrix}
		\mathbf{Q}_{\mathrm{n},\head}^{H},\hspace{2pt}\mathbf{q}_{\mathrm{n},\eMicInd}^{\ast}
	\end{bmatrix}
	\begin{bmatrix}
		\vectorADomwhiteCMA \\ A_{\eMicInd}^{\mathrm{w}}\left(\thetaDom\right)
	\end{bmatrix} = \mathbf{0}_{M+1 \times 1}\,,
	\label{eq:MUSICorthogonality}
\end{align}
with $\vectorADomwhiteCMA = \selectionMatrixHA\vectorADomwhite$ and $A_{\eMicInd}^{\mathrm{w}}\left(\thetaDom\right) = \mathbf{e}_{\eMicInd}^{T}\vectorADomwhite$ and with $\left(\cdot\right)^{\ast}$ denoting the element-wise complex conjugation operation. We stress that the orthogonality relation in \eqref{eq:MUSICorthogonality} relates \textit{all} pre-whitened ATFs with the \textit{full} noise subspace. Please note that in general $\mathbf{Q}_{\mathrm{n},\head}^{H}\vectorADomwhiteCMA\neq \mathbf{0}_{M \times 1}$.

We propose to exploit the orthogonality relation in \eqref{eq:MUSICorthogonality} for the computation of the pre-whitened ATF $A_{\eMicInd}^{\mathrm{w}}\left(\thetaDom\right)$ given $\mathbf{Q}_{\mathrm{n},\head}$, $\mathbf{q}_{\mathrm{n},\eMicInd}$ and $\vectorADomwhiteCMA$ using a least-squares optimization problem, i.e.,
\begin{equation}
	\label{eq:novelty_pt1}
	\alpha_{\mathrm{opt}} = \underset{\alpha}{\mathrm{argmin}}\,\big\|\mathbf{Q}_{\mathrm{n},\head}^{-H}\mathbf{q}_{\mathrm{n},\eMicInd}^{\ast} - \alpha\,\vectorADomwhiteCMA\big\|_{2}^{2}\,,
\end{equation}
with $\alpha = -\frac{1}{A_{\eMicInd}^{\mathrm{w}}\left(\thetaDom\right)}$. The solution to \eqref{eq:novelty_pt1} is given by
\begin{equation}
	\boxed{\alpha_{\mathrm{opt}} = \frac{\vectorADomHermwhiteCMA\mathbf{Q}_{\mathrm{n},\head}^{-H}\mathbf{q}_{\mathrm{n},\eMicInd}^{\ast}}{\big\|\vectorADomwhiteCMA\big\|_{2}^{2}}\,\Rightarrow \quad A_{\eMicInd}^{\mathrm{w}}\left(\thetaDom\right) = -1/\alpha_{\mathrm{opt}}}
	\label{eq:novelty_pt2}
\end{equation}
Thus, given an ATF vector $\vectorADomwhite = \begin{bmatrix}
	\vectorADomwhiteCMA,\hspace{2pt}A_{\eMicInd}^{\mathrm{w}}\left(\thetaDom\right)
\end{bmatrix}^{T}$ with unknown element $A_{\eMicInd}^{\mathrm{w}}\left(\thetaDom\right)$, using \eqref{eq:novelty_pt2} one can complete the vector $\vectorADomwhiteCMA$ by exploiting the orthogonality relation between the pre-whitened ATFs and the noise subspace.

Based on the least squares solution in \eqref{eq:novelty_pt2}, we propose the following completed set of pre-whitened prototype ATF vectors
\begin{equation}
	\label{eq:mainProp}
	\boxed{\bar{\mathbf{a}}^{\mathrm{w}}_{\mathrm{completed}}\left(k,l,\theta_{i}\right) = \begin{bmatrix}
			\bar{\mathbf{a}}_{\head}^{\mathrm{w}}\left(k,l,\theta_{i}\right)\\
			-\frac{\big\|\bar{\mathbf{a}}^{\mathrm{w}}_{\head}\left(k,l,\theta_{i}\right)\big\|_{2}^{2}}{\left(\bar{\mathbf{a}}^{\mathrm{w}}_{\head}\right)^{H}\left(k,l,\theta_{i}\right)\hat{\mathbf{Q}}_{\mathrm{n},\head}^{-H}\left(k,l\right)\hat{\mathbf{q}}_{\mathrm{n},\eMicInd}^{\ast}\left(k,l\right)}\end{bmatrix}}
\end{equation}
Based on \eqref{eq:CW} and \eqref{eq:mainProp}, we propose the following completed set of prototype RTF vectors
\begin{equation}
	\label{eq:mainProp2}
	\boxed{
	\bar{\mathbf{g}}_{\mathrm{completed}}\left(k,l,\theta_{i}\right) = \frac{\hatphiU^{1/2}\left(k,l\right)\hspace{2pt}\bar{\mathbf{a}}_{\mathrm{completed}}^{\mathrm{w}}\left(k,l,\theta_{i}\right)}{\mathbf{e}_{1}^{T}\hatphiU^{1/2}\left(k,l\right)\hspace{2pt}\bar{\mathbf{a}}_{\mathrm{completed}}^{\mathrm{w}}\left(k,l,\theta_{i}\right)}}
\end{equation} 
Using the completed sets of prototype transfer function vectors in \eqref{eq:mainProp} and \eqref{eq:mainProp2}, allows to construct frequency-dependent SPS according to \eqref{eq:MUSIC} and \eqref{eq:RTFprototypeMatching} when used with partially calibrated arrays. Fig. 2 summarizes the novel DOA estimation methods.

\section{Experimental results}
\label{sec:experiments}
For acoustic scenarios with two static speakers in multiple reverberant environments with diffuse-like babble noise, in this section we compare the DOA estimation performance with MUSIC and the RTF vector matching method when using the incomplete sets $\bar{\mathbf{a}}_{\head}^{\mathrm{w}}\left(k,l,\theta_{i}\right)$ and $\bar{\mathbf{g}}_{\head}\left(k,\theta_{i}\right)$ and the completed sets $\bar{\mathbf{a}}_{\mathrm{completed}}^{\mathrm{w}}\left(k,l,\theta_{i}\right)$ in \eqref{eq:mainProp} and $\bar{\mathbf{g}}_{\mathrm{completed}}\left(k,l,\theta_{i}\right)$ in \eqref{eq:mainProp2}. In Section \ref{ssec:experiments__setup}, we describe the experimental setup and implementations details. In Section \ref{ssec:experiments__res}, we present and discuss the results.

\subsection{Experimental setup and implementation details}
\label{ssec:experiments__setup}
For the experiments, we consider separate recordings of speech and diffuse-like babble noise from the BRUDEX database \cite{Fejgin2023ITG}. The signals have been recorded in a laboratory at the University of Oldenburg with dimensions of about $(7\times6\times2.7)\,\mathrm{m}^{3}$ with binaural hearing aids ($M=4$) on a dummy head and an eMic placed at 36 possible locations which are uniformly distributed in the laboratory. We consider three reverberation environments ('low', 'medium', and 'high'), corresponding to median reverberation times $\mathrm{T}_{60}\approx\left[310,510,1300\right]\hspace{0.5pt}\mathrm{ms}$. Excluding co-located speakers, we consider a female and a male speaker ($J=2$, both constantly active, duration = $5\hspace{0.5pt}\mathrm{s}$) located at 132 possible two-speaker DOA combinations in the range $\left[-150:30:180\right]\,^{\circ}$ at a distance of approximately $2\,\mathrm{m}$ relative to the dummy head. For both speakers we consider equal average broadband speech power across all signals of the microphones of the hearing aids. The noise component is added to the reverberant speech component after scaling the broadband noise power across all signals of the microphones of the hearing aids to signal-to-noise ratios (SNR) in the range $\left[-5:5:20\right]\hspace{0.5pt}\mathrm{dB}$. All microphone signals are assumed to be exchanged without quantization errors and are assumed to be synchronized.

To assess the benefit of DOA estimation with the completed sets  $\bar{\mathbf{a}}_{\mathrm{completed}}^{\mathrm{w}}\left(k,l,\theta_{i}\right)$ and $\bar{\mathbf{g}}_{\mathrm{completed}}\left(k,l,\theta_{i}\right)$ over the incomplete sets $\bar{\mathbf{a}}_{\head}^{\mathrm{w}}\left(k,l,\theta_{i}\right)$ and $\bar{\mathbf{g}}_{\head}\left(k,\theta_{i}\right)$, we compare the following conditions for MUSIC and the RTF vector matching method:
\begin{itemize}
	\setlength{\itemindent}{0.4cm}
	\item[H/H:] EVD of $\hatphiYwhiteHA$ (no eMic) and proto\-type matching with $\bar{\mathbf{a}}_{\head}^{\mathrm{w}}\left(k,l,\theta_{i}\right)$ and $\bar{\mathbf{g}}_{\head}\left(k,\theta_{i}\right)$ (no eMic),
	\setlength{\itemindent}{0.75cm}
	\item[H+E/H:] EVD of $\hatphiYwhite$ (with eMic) and proto\-type matching with $\bar{\mathbf{a}}_{\head}^{\mathrm{w}}\left(k,l,\theta_{i}\right)$ and $\bar{\mathbf{g}}_{\head}\left(k,\theta_{i}\right)$ (no eMic),
	\setlength{\itemindent}{1.1cm}
	\item[H+E/H+E:] EVD of $\hatphiYwhite$ (with eMic) and proto\-type matching with $\bar{\mathbf{a}}_{\mathrm{completed}}^{\mathrm{w}}\left(k,l,\theta_{i}\right)$ and $\bar{\mathbf{g}}_{\mathrm{completed}}\left(k,l,\theta_{i}\right)$ (with eMic).
\end{itemize}

All microphone signals are downsampled to $16\,\mathrm{kHz}$. The algorithms are implemented within an STFT framework with $32\,\mathrm{ms}$ square-root Hann windows with $50\,\%$ overlap. We estimate the covariance matrices $\hatphiY\left(k,l\right)$ and $\hatphiU\left(k,l\right)$ for each TF bin using a first order recursion during speech-and-noise periods and noise-only periods, respectively, as
\begin{align}
	\hatphiY\left(k,l\right) &= \alpha_{\mathrm{y}}\hatphiY\left(k,l-1\right) + \left(1-\alpha_{\mathrm{y}}\right) \mathbf{y}\left(k,l\right)\mathbf{y}^{H}\left(k,l\right)\,,\\
	\hatphiU\left(k,l\right) &= \alpha_{\mathrm{u}}\hatphiU\left(k,l-1\right) + \left(1-\alpha_{\mathrm{u}}\right) \mathbf{y}\left(k,l\right)\mathbf{y}^{H}\left(k,l\right)\,,
\end{align}
with smoothing factors $\alpha_{\mathrm{y}}$ and $\alpha_{\mathrm{u}}$ corresponding to time constants of $250\,\mathrm{ms}$ and $500\,\mathrm{ms}$, respectively. To discriminate speech-and-noise periods from noise-only periods, speech presence pro\-ba\-bilities \cite{Gerkmann2012} are estimated using the hearing aid microphone signals, averaged and thresholded. Based on the results reported in \cite{Fejgin2024} for the selection of the frequency subset $\mathcal{K}$, we set $\mathrm{CDR}_{\mathrm{thresh}} = -3\,\mathrm{dB}$ for MUSIC and $\mathrm{CDR}_{\mathrm{thresh}} = -5\,\mathrm{dB}$ for the RTF vector matching method. The indicator function $\indicatorFunctionKL$ is estimated as described in \cite{Fejgin2024}.

The set of anechoic prototype ATF vectors $\bar{\mathbf{a}}_{\head}\left(k,\theta_{i}\right)$ is obtained from measured anechoic binaural room impulse responses \cite{Kayser2009} with an angular resolution of $5\,^{\circ}$ in the range $\left[-180:5:175\right]^{\circ}$ ($I=72$).

We assess the DOA estimation performance using the following definition of accuracy:
\begin{equation}
	\mathrm{ACC} = \frac{1}{J\,L}\sum_{l=1}^{L}j_{\mathrm{correct}}\left(l\right)\,,
\end{equation}
where $j_{\mathrm{correct}}$ denotes the number of speakers for which the DOA is estimated within $\pm 5^{\circ}$ correctly. For both DOA estimation methods, we average the accuracies over all acoustic scenarios, i.e., DOA combinations, SNRs, and reverberation conditions.

\subsection{Results}
\label{ssec:experiments__res}
\begin{figure}[t]
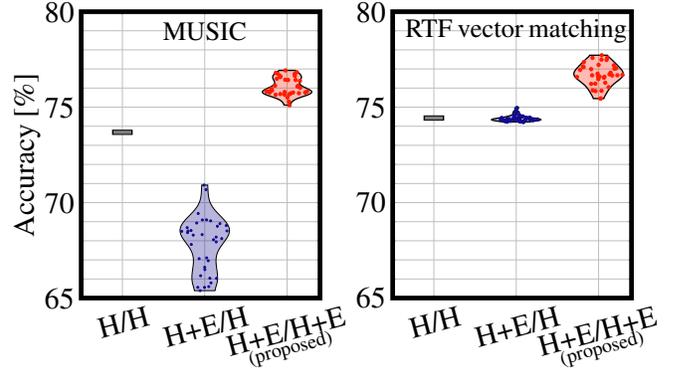

	\addtocounter{figure}{1}
	\centering
	\input{resMUSIC_violinPlot.tex}\input{resRTF_violinPlot.tex}
	\vspace{-1mm}
	\caption{Average localization accuracy of DOA estimation using MUSIC (left) and the RTF vector matching method (right) with the conditions H/H (no eMic), H+E/H (eMic included only in the EVD of $\hatphiYwhite$ but not in the prototype matching), and H+E/H+E (eMic included in the EVD of $\hatphiYwhite$ and in the prototype matching).\label{fig:results}} 
	\vskip-3.4mm
\end{figure}
Fig. 3 depicts the average localization accuracy for the investigated conditions for the two DOA estimation methods. The horizontal line in the H/H condition (no eMic) shows the average localization accuracy when considering only the  hearing aid microphone signals. Considering in addition to these signals also the eMic signal, the violin plots in the H+E/H condition (eMic signal included only in the EVD of $\hatphiYwhite$ but not in the sets of prototype transfer function vectors) and in the H+E/H+E condition (eMic signal included in the EVD of $\hatphiYwhite$ and in the sets of prototype transfer function vectors) show the distribution of average localization accuracy due to different locations of the eMic. First, it can be observed that for both DOA estimation methods the performance with the H/H condition is similar (about $74 \%$). We interpret this observation as a consequence that the signal subspace and the noise subspace are similarly meaningful for DOA estimation since they are obtained from the same covariance matrix $\hatphiYwhiteHA$. Second, it can be observed that for the RTF vector matching method there is only a minor performance difference between the H/H and the H+E/H condition. We interpret this result, which is in line with the results reported in \cite{Fejgin2021} and \cite{Fejgin2023}, as a consequence of the parallelity of the vectors $\mathcal{P}\{\PSDs\vectorADomwhiteCMA\vectorADomHermwhiteCMA\}$ and $\selectionMatrixHA\mathcal{P}\{\PSDs\vectorADomwhite\vectorADomHermwhite\}$. For MUSIC, however, there is a large performance difference between the H/H and the H+E/H condition. We interpret the higher performance at the H/H condition compared to the H+E/H condition as a consequence of the orthogonality relation in \eqref{eq:MUSICorthogonality}, which holds for all signals and not just a subset of signals. Also note that in MUSIC the condition H+E/H can be understood as setting $A_{\eMicInd}^{\mathrm{w}}\left(\thetaDom\right)$ to $0$, which clearly deviates from the optimal solution in \eqref{eq:novelty_pt2}. Third, considering the H+E/H+E condition (corresponding to the proposed completion procedure), the results clearly show that for both methods DOAs can be estimated more accurately than with the H/H and H+E/H conditions. This result together with those reported in \cite{Fejgin2023} further supports the advantage in exploiting eMic signals for the construction of spatial spectra. Fourth, for both DOA estimation methods the results from the H+E/H+E condition show the highest average localization accuracies for \textit{all} eMic locations. Based on these results, the potential and robustness to the location of the eMic of the proposed procedure for completing sets of prototype transfer functions is clearly demonstrated.

\section{Conclusions}
\label{sec:conclusions}
In this paper we estimated DOAs of multiple speakers with partially calibrated microphone arrays, composed of a calibrated binaural hearing aid and a non-calibrated external microphone at an unknown location. We proposed an optimal procedure in the least-squares sense that exploits the external microphone for the completion of sets of prototype transfer function vectors. We compared DOA estimation with the incomplete and completed sets of prototype transfer function vectors for the subspace-based MUSIC and the RTF vector matching method. Experimental results with two speakers in multiple reverberant environments with diffuse-like noise from the BRUDEX database clearly demonstrate that DOAs can be estimated more accurately with the proposed completed sets of prototype transfer function vectors than with incomplete sets. Moreover, we showed that the procedure is robust to the location of the external microphone.

\bibliographystyle{IEEEtran}
\bibliography{myRefs.bib}

\end{document}